\documentclass[prd,superscriptaddress,showpacs,preprintnumbers,amsmath,amssymb]{revtex4}
\usepackage{amsmath}
\usepackage{amstext}
\usepackage{amsopn}
\usepackage{amsfonts}
\usepackage{amssymb}
\usepackage{bbm}
\usepackage{accents}
\usepackage{empheq}
\usepackage{graphicx}
\usepackage{epsf}
\usepackage{graphics}
\usepackage[latin1]{inputenc}

\begin{document}

\title{Warming up D3 brane motion in the background of D5 brane and Inflation}
\author{Anindita Bhattacharjee }
\author{Atri Deshamukhya\footnote{atri.deshamukhya@gmail.com}}
\affiliation{Department of Physics, Assam University, Silchar, Assam 788011, India}

\begin{abstract}
The position of a mobile D-3 brane moving towards a  stack of localized D-5 branes has been studied as a candidate driving inflation in the warm inflationary scenario. We compare the results obtained by considering the dissipation parameter $\Gamma$ as an arbitrary function of only the inflaton field and a particular form derived by Bastro-Gil et al \cite{Berera8}. We find that the observables remain well within the recent observational constraint for a wide range  of model parameters for the first case whereas the spectral index in the later case is always predicted blue, other cosmological observables remaining well within bound for a wider range of parameters though. We also discuss the non-gaussianity generated during inflation in this model.
\end{abstract}


\maketitle

\section{Introduction}

In past few years there have been many attempts to build cosmological inflation models in the framework of string theory. One of the popular class of models is based on a scenario with a brane moving towards a localized antibrane, where the position of the mobile brane plays the role of the inflaton field \cite{Dvali,Burgess,Garcia,Har,KKLMT,Bau}. In another approach, for inflation model building the open string tachyon field described by a non-standard action \cite{Panda0}, which lives in a world volume of a non-BPS brane, has been used as the inflaton field \cite{Panda1}. It was observed that the same non-standard tachyon effective action also describes the dynamics of a mobile BPS brane in the background of a stack of BPS branes \cite{Kuta}.Inflationary models have been constructed in this set-up \cite{Geometric}. In particular, an inflation model, where a D3 brane moves in the background of a stack of k- coincident D5 branes was investigated in \cite{Panda2}. It was noted that the inflation ends much before the mobile brane comes to a distance of the order of string length scale, from the stack of localized branes. Thus, the tachyon field, which would have been excited, when they are close enough, does not play any role in driving the inflation.

However, from thermodynamic viewpoint, there are two dynamical realizations of inflation.
The earlier version is the standard inflation scenario (also known as supercooled inflation) where radiation is red-shifted during expansion and leads to a vacuum dominated universe. This gives an isentropic perspective of the inflation paradigm where the universe expands almost exponentially in inflation phase and as a result  its temperature decreases rapidly. Radiation is being introduced through a reheating period after the end of inflation. The fluctuations during this inflation phase are zero-point ground state fluctuations and inflaton field evolution is governed by the ground state evolution equation. There are no thermal perturbations and therefore, density perturbations here are only adiabatic in nature . In these type of models  expansion and reheating are two distinguished phases. Also, energy transfer from potential energy to radiation remains a nontrivial aspect of supercooled inflation \cite{Nojari}. Cold Inflation, in fact, is an idealized situation where the dynamics reduces to the classical evolution of the scalar inflaton field with vacuum quantum fluctuations superposed on this background field.

In contrast to the cold inflationary picture, warm inflation, the other thermodynamic alternative,  presents
the attractive feature of avoiding the reheating period altogether \cite{Berera}. In such type of models dissipative effects are important during the inflationary era, so that radiation production can occur together with the inflationary expansion.  The dissipating effect is the result of the friction arising from the scalar field dissipating into a thermal bath via its interaction with other fields during the period of inflation \cite{Herera1}. Phenomenologically in the interaction process, the inflaton decays in to some other fields and the decay of the scalar field can be  described by means of an interaction Lagrangian. From the point of view of statistical mechanics, the interaction between quantum fields and a thermal bath could be demonstrated by a general fluctuation-dissipation relation \cite{Weiss}. Warm inflation shows how thermal fluctuations may play a dominant role in producing the initial fluctuations necessary for the formation of Large- Scale Structures. Here, the density fluctuations arise from thermal rather than quantum fluctuations as it happens in supercooled inflation \cite{Berera2}. These fluctuations have their origin in the hot radiation and their influence on the inflaton scalar field is  introduced through a friction term in its equation of motion  \cite{Berera3}.

Warm inflation was criticized on the basis that the inflaton field cannot decay during the slow-roll phase \cite{Linde} of inflationary expansion. However, it can be shown that the inflaton field can indeed decay during the slow-roll phase (see \cite{Berera5} and references therein) whereby placing the concept of warm inflationary paradigm on solid theoretical ground. Over the years, the theory of the dissipation coefficient has met immense success in the high temperature regime with the condition $T_{\gamma} > m_{\chi}$ where, $\chi$ is the field interacting with the inflaton and $T_{\gamma}$ is the radiation temperature. The warm Inflationary scenario in low temperature regime is also of great interest now a days which involves a two-stage decay procedure $\phi \rightarrow \chi \rightarrow yy$ where, $\chi$ is the heavy intermediate field and $y$ is the finally decaying weak field \cite{Moss*}. As the decay processes in both the regimes are not same, so, the density perturbation can be expressed in terms of the dissipation coefficients with different temperature dependence- which will certainly give rise to different observational consequences \cite {Hall*}. Recently Bastero-Gil et al has obtained an expression for the associated dissipation coefficient in supersymmetric models\cite{Berera8}in low temperature regime.In warm inflationary scenario, in presence of radiation in early universe, the idealization in general made is of a perfect fluid whereas some deviations might be there from this limit leading to viscous dissipation and corresponding noise forces which might have observational consequences \cite{Berera11}.

Warm inflation ends when the universe heats up to become radiation dominated. At this epoch, the universe stops inflating and smoothly enters into a radiation dominated Big Bang phase \cite{Berera6}. The matter components of the universe are created by the decay of either the dominant radiation field or the remaining inflationary field \cite{Berera7}.

Warm inflation has been studied in the context of tachyon cosmology \cite{Herera, Atri}, brane-antibrane scenario \cite {Herera1, Rosa}, geometric-tachyon driven case \cite{Atri1}. In the present piece of work, we revisit the inflationary scenario driven by the radion field of separation between a $D3$ and a stack of $D5$ branes in presence of thermal bath.In Section.II we describe the warm inflationary scenario in the present context considering the dissipation coefficient an arbitrary function of the inflaton field. In section.III we consider the form of the dissipation coefficient derived by Bastro-Gil et al and a cosmological re-analysis has been done. Section.IV is dedicated for conclusion and discussion.

\section{Warm inflation driven by the Radion Field between  $D3$ brane and stake of $D5$ branes where the dissipation coefficient $\Gamma$ is arbitrary function of the inflaton field}

\subsection{Review of Formalism}
Dynamics of a warm-inflationary model where the inflaton field can be described by non-standard tachyonic action for flat FRW metric are described by the equations \cite{Herera}:
\begin{equation}
H^{2}=\frac{1}{3M_{p}^{2}}\frac{V}{\sqrt{1-\dot{T}^{2}}},
\end{equation}
\begin{equation}
\frac{\ddot{T}}{\sqrt{1-\dot{T}^{2}}}+3H\dot{T}+\frac{\Gamma \dot{T}}{V}\sqrt{1-\dot{T}^{2}}=-\frac{V_{,T}}{V},
\end{equation}

 and
\begin{equation}
{\dot{\rho}}_{\gamma}+4H\rho_{\gamma}=  \Gamma {\dot{T}}^2.
\end{equation}
where, $M_{P}$ is the reduced Planck mass and  $V_{,{T}}\equiv \partial V/\partial T$ and  over dots represent derivative with respect to real time.
$\Gamma$ is the dissipation co-efficient responsible for the decay of scalar field into radiation during inflationary epoch and $\rho_{\gamma}$ is the energy density due to radiation. $\Gamma$ must satisfy $ \Gamma > 0$ by the second law of thermodynamics. In the anlysis of this section, we consider $\Gamma$ as a function of only the scalar field only though in principle it should be a function of both scalar field as well as temperature.

Now to have an inflationary scenario, the necessary conditions are $\rho_{T}~ \sim ~ V$ and $\rho_{T}~ > ~\rho_\gamma$. Also, the slow-roll approximation demands that ${\dot{T}}^2\ll 1$ and $\ddot{T}\ll (3H+\frac{\Gamma}{V})\dot{T}$. Under these conditions, the evolution equations can be written as \cite{Herera}:
\begin{eqnarray}
H^2~&=&~\frac{V}{3 M_P^2}\\
3H(1+r){\dot{T}}~&=&~-\frac{V,_{T}}{V}.
\end{eqnarray}
where, we have defined a dimensionless parameter $r$ as $r\equiv\frac{\Gamma}{3HV}$.

 In addition, to have a quasi-stable radiation production during inflationary epoch, the necessary conditions are $\dot{\rho_\gamma} \ll 4H\rho_\gamma$ and $\dot{\rho_\gamma}\ll \Gamma{\dot{T}}^2$. With slow-roll conditions, we then have
\begin{equation}
\rho_\gamma~=~\frac{\Gamma {\dot{T}}^2}{4H}.
\end{equation}
On the other hand $\rho_\gamma$ can be written as $\rho_\gamma \equiv \sigma {T_\gamma}^4$ where $\sigma$ is the Stephan-Boltzmann constant and $T_\gamma$ is temperature of the thermal bath.

By using Eq.5 and Eq.6 with the definition of $r$,we get
\begin{equation}
\rho_\gamma\equiv \sigma {T_\gamma}^4= \frac{r~M_P^2}{4 (1+r)^2}\left(\frac{V,_{T}}{V}\right)^2
\end{equation}

The combination $\sigma {T_\gamma}^4$ can be chosen as a dimensionfull parameter $d$ and all the observable quantities in a warm tachyonic inflationary scenario can be expressed in terms of this parameter.

The dimensionless slow-roll parameters in this model are expressible as :
\begin{eqnarray}
\varepsilon&\equiv& -\frac{\dot{H}}{H^2}={\frac{M_P^2}{2 (1+r)V}}\left(\frac{V,_T}{V}\right)^2\\
\eta&\equiv& -{\frac{\ddot{H}}{H\dot{H}}}={\frac{M_P^2}{(1+r)
V}}\left[\frac{V,_{TT}}{V}-{{\frac{1}{2}}\left(\frac{V,_T}{V}\right)^2}\right].
\end{eqnarray}
With $r=0$, the above equations reproduces the usual cold-inflation expressions. The inflation ends when either of the parameters $\varepsilon$ or $\eta$ goes to one ( whichever is early).
Number of e-folds from an arbitrary field value to the end of inflation is given by
\begin{equation}
N(T)~=~-\frac{1}{M_P^2}\int^{T_e}_T {\frac{V^2}{V,_{T'}}}(1+r)~dT'.
\end{equation}
where $T_e$ is the field magnitude at the end of inflation.
For perturbations, both scalar and tensor, we note that in the case of scalar perturbations the scalar and the radiation fields are interacting.
Therefore, isocurvature (or entropy) perturbations are generated besides of the adiabatic ones. This occurs because warm inflation can be considered as an inflationary model with two basics fields. In this context dissipative effects can produce a variety of spectral,
ranging between red and blue, and thus producing the running blue to red spectral\cite{Herera1}.

At the high dissipation regime, the density perturbations is expressed as \cite{Herera}:
\begin{equation}
\delta_{H}=\frac{16~\pi}{5}\frac{M_{p}^{2} \exp(-\bar{\zeta}(T)}{(\ln(V)),_{T}} \delta T,
\end{equation}
where,
\begin{eqnarray}
\bar{\zeta}(T)&=&-\int\left[\frac{1}{3Hr}\left(\frac{\Gamma}{V}\right),_{T}
+\frac{9}{8}\left(1-\frac{(\ln(\Gamma)),_{T}(\ln(V)),_{T}}{36 H^{2}r}\right) (\ln(V)),_{T}\right] dT\\
&=&-\int\left[2\frac{V,_{TT}}{V,_{T}}-\frac{3V,_{T}}{8V}-\frac{3d~V,_{TT}}{4VV,_{T}}+\frac{3d~V,_T}{16V^{2}}\right] dT
\end{eqnarray}

In terms of the slow-roll parameter $\varepsilon$, the quantity $\delta^{2}_{H}$ can be expressed as :

\begin{equation}
\delta^{2}_{H}= \frac{\sqrt{3}}{75~\pi^{2}}\exp \left[-2 \bar{\zeta}(T)\right] \left[\left(\frac{1}{\varepsilon}\right)^3 \left(\frac{9 M_{P}^{4}}{2 r^{2} \sigma V}\right)\right]^{\frac{1}{4}}
\end{equation}
which gives rise to expressions for various cosmological observables. For example, the spectral index defined by
$n_{s} \equiv 1+\frac{d~ln\delta_H^2}{d~ln k}$
becomes
\begin{equation}
 n_s =1-\left[\frac{3\eta}{2}
+\epsilon\left[\frac{2V}{V,_{T}}(2\bar{\zeta}(T)-\frac{r_{T}}{4r})-\frac{5}{2}\right]\right]
\end{equation}
and the running spectral index is found to be
\begin{eqnarray}
 \alpha_{s}&\equiv&\frac{d~n_s}{d~ln~k}\\ \nonumber
 &=& -\frac{2V\epsilon}{V,_{T}}\left[\frac{2\eta,_{T}}{2}
 +\frac{\epsilon_{T}}{\varepsilon}\left[n_{s}-1+\frac{3\eta}{2}\right]
 +2\epsilon\left[\left(\frac{V}{V,_{T}}\right),_{T}\left(2\bar{\zeta},_{T}
 -\frac{(\ln(r)),_{T}}{4}\right)
 +\left(\frac{V}{V,_{T}}\right)\left(2\bar{\zeta},_{TT}-\frac{(\ln(r)),_{TT}}{4}\right)\right]\right]
\end{eqnarray}

Similarly, the Power spectrum defined by
\begin{equation}
 P \equiv \frac{25}{4}\delta_{H}^{2}
\end{equation}
is expressible as
\begin{equation}
P=\frac{1}{4\pi^{2}}\left[\frac{1}{\sigma d}\frac{V^{6}}{(V,_{T})^{4}}\right]^{\frac{1}{4}}\exp(-2\bar{\zeta}(T))
\end{equation}
and the tensor to scalar ratio is given by
\begin{equation}
R\equiv\left[\frac{A_{g}^{2}}{P}\right]_{k=k_{0}}
\end{equation}
where $A_{g}^{2}$ is called the tensor spectrum which is expressed as
\begin{equation}
 A_{g}^{2}=\frac{H^{2}}{2\pi^{2}M_{p}^{2}}\left[\coth\left[\frac{k}{2T_{\gamma}}\right]\right]_{k=k_{0}}.
\end{equation}

\subsection{Warm Inflation driven by the Radion field}

When we consider the motion of a $D3$ brane in the background of $k$ coincident $D5$ brane, the scalar field describing the distance between them becomes tachyonic and can be expressed as
\begin{equation}
T(R)=\sqrt{L^2+R^2}+\frac{1}{2}~L~\ln\frac{\sqrt{L^2+R^2}-L}{\sqrt{L^2+R^2+L}},
\end{equation}
where, $R$ is the distance between the moving $D3$ and $k$ number of static $D5$ branes. $L$ is defined as $L=\sqrt{k~g_{s}l_{s}^2}$, $g_s$being the string coupling and $l_s$ being the string length scale.

Now, for this tachyonic field T, the potential function can be written as \cite{Panda2}
\begin{equation}
V(T)=\tau_{3}\frac{x}{\sqrt{x^2+1}}= \tau_{3} V(x),
\end{equation}
where $\tau_3$ is the tension on the branes and $x=\frac{R}{L}$ and $x$ is related to $T$as
\begin{equation}
\Rightarrow\frac{dT}{dx}=\frac{L}{V(x)}.
\end{equation}

Thus, in terms of the dimensionless parameter $x$  the tachyonic field $T$ can be expressed as :
\begin{equation}
T=L \left(\sqrt{1+x^2}+\frac{1}{2} \log  \left( \frac{\sqrt{1+x^2}-1}{\sqrt{1+x^2}+1} \right)\right)
\end{equation}

In terms of the parameters $r$ and $d$  defined by $r=\frac{\Gamma}{3HV}=\frac{M_{p}^{2}(V_{,T})^{2}}{4dV^{2}}$ and $d=\sigma(T_{\gamma})^{4}$, the number of e-folds from some initial time to the end of inflation is found to be
\begin{eqnarray}
N\equiv \int_{t_i}^{t_{f}} H dt &=& -{\int}_{T}^{T_f}\frac{3H^{2}r V(T)}{V,_{T}}dT\\ \nonumber
                 &=& p[V(x)-V(x_{f})]\\  \nonumber
                 &=& p\left[\frac{x}{\sqrt{x^2+1}}-\frac{x_{f}}{\sqrt{x_{f}^2+1}}\right].
\end{eqnarray}
where $p$ is another parameter which is defined as $p =\frac{\tau_{3}}{4d}$.

Accordingly, the slow-roll parameter $\varepsilon$ and $\eta$ takes the form
\begin{equation}
\varepsilon=\frac{2d}{\tau_{3}V(x)}=\frac{1}{2pV(x)}
\end{equation}
and
\begin{equation}
\eta =\frac{1}{2pV(x)}(1-6 x^2)= \varepsilon (1-6 x^2)
\end{equation}

From Eq.26 and Eq.27, it is evident that the epsilon approaches 1 faster as the field rolls from higher to lower value for any value of the parameter $p$ so long as $x\leqslant1$ i.e $R\leqslant L$. Hence the end of inflation is marked by the condition :
\begin{equation}
\frac{1}{2pV(x)}=1,
\end{equation}
which in turn gives
\begin{equation}
x_{f}=\frac{1}{\sqrt{4p^2-1}}.
\end{equation}
Putting the value of $x_{f}$ in the expression of $N$, $x$ becomes
\begin{equation}
x=\frac{1}{\sqrt{\frac{p^2}{(N+\frac{1}{2})^2}-1}}.
\end{equation}

Thus, we see that the parameter $x$ can be expressed in terms of $p$ and $N$. It may also be noted that $R\leqslant L$ imposes a constrain $p\geqslant 86 $.

After the introduction of all these parameters,we are now ready to compute all the observable which arise from perturbation spectrum of CMB.

\subsubsection{Perturbational analysis}

\begin{flushleft}
\bf{(a) Spectral Index : }
\end{flushleft}
The spectral index in terms of model parameters may be expressed as
\begin{eqnarray}
n_{s} &=&1-\left[\frac{3\eta}{2}+\epsilon\left[\frac{2V}{V,_{T}}(2\bar{\zeta}(T)-\frac{r_{T}}{4r})-\frac{5}{2}\right]\right]\\ \nonumber
      &=&~\frac{3d^{2}}{2V^{2}}+\frac{3d}{V}+1-\frac{6d^{2}V,_{TT}}{V(V,_{T})^{2}} +\frac{12V,_{TT}d}{(V,_T)^{2}}\\ \nonumber
      &=& 1+\left(\frac{3}{32p^2 V(x)}+\frac{3}{4 p}\right)\frac{1}{V(x)}-\frac{3}{8 p^2}\left(\frac{1}{V(x)^2}+\frac{V''(x)}{V(x) V'(x)^2}\right)+\frac{3}{p}\left(\frac{1}{V(x)}+\frac{V''(x)}{ V'(x)^2}\right)\\ \nonumber
      &=&1+\frac{3[-3+x(40p\sqrt{1+x^2}
      +3x[3+4x(x-8p\sqrt{1+x^2})])]}{32p^2x^2}
\end{eqnarray}

 As, $x$ is a function of both $p$ and $N$, for a fixed $N$, the spectral index can be expressed in terms of the parameter $p$ and $N$.

 \begin{figure}[htbp]
  \centering
  \includegraphics[width=8cm]{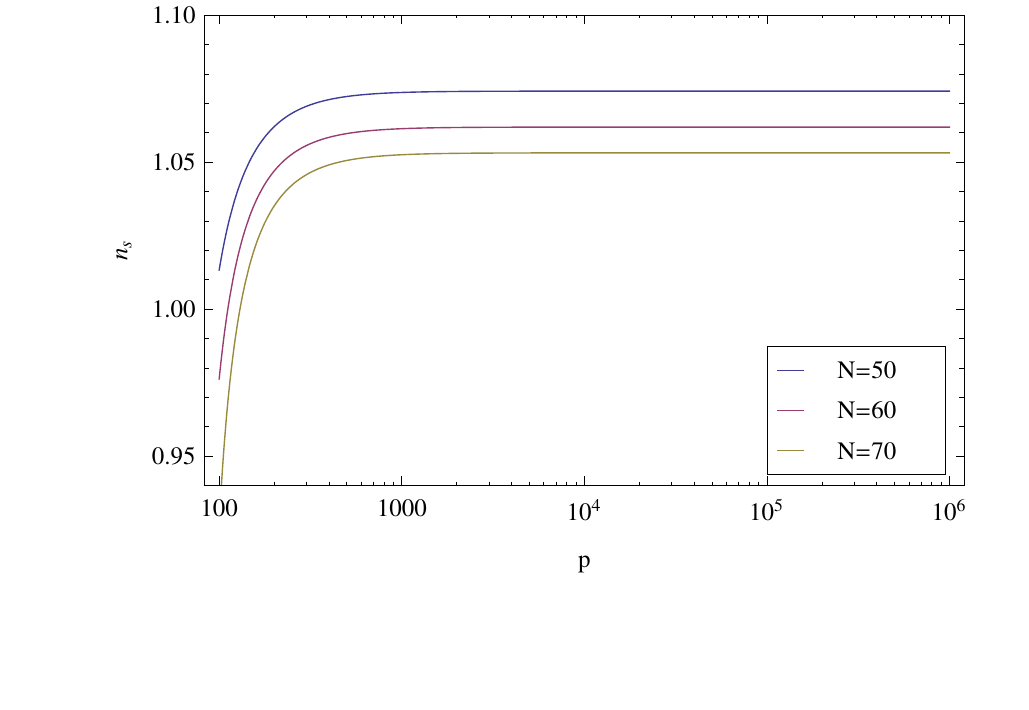}\\
  \caption{Variation of Spectral Index with $p$ for N=70(brown),60(red) and 50(blue) respectively}
\end{figure}

 In Fig.1, variation of spectral index with respect to $p$ for three different $N$ values have been shown. It can be seen that for a wide range of parameter $p$, the value of $n_{s}$ lies well within the limit as predicted by WMAP9 \cite{WMAP9} as well as PLANCK \cite{PLANK}. For example, When p=100 and N=60, $n_{s}$ becomes 0.976197 which well within the bound of recent observation.

 \begin{flushleft}
\bf{(b) Running spectral index :}
\end{flushleft}
 The running of the spectral index is another observable associated with the spectral index and it can be defined as
 \begin{eqnarray}
 \alpha_{s} &=&-\frac{2V\varepsilon}{V,_{T}}\left[\frac{2\eta,_{T}}{2}
                +\frac{\varepsilon_{T}}{\varepsilon}\left[n_{s}-1+\frac{3\eta}{2}\right]
                +2\varepsilon\left[\left(\frac{V}{V,_{T}}\right),_{T}\left(2\bar{\zeta},_{T}
                -\frac{(\ln(r)),_{T}}{4}\right)
                +\left(\frac{V}{V,_{T}}\right)\left(2\bar{\zeta},_{TT}-\frac{(\ln(r)),_{TT}}{4}\right)\right]\right]\\ \nonumber
            &=&-\frac{12d^{2}}{V^{3}(V,_T)^{4}}[(d+V)(V,_T)^{4}-2dV(V,_T)^{2}V,_{TT}+4V^{2}[-d+2V](V,_{TT})^{2}
                +2[d-2V]V^{2}V,_T~V,_{TTT}]\\ \nonumber
            &=& - (\frac{3}{16p^3V(x)^3}+\frac{3}{4p^2V(x)^2}
 -\frac{3[V(x)V'(x)^2+V(x)^2V''(x)]}{8p^3V(x)^4V'(x)^2}
 -\frac{3[V(x)V'(x)^2+V(x)^2V''(x)]^2}{4p^3V(x)^5V'(x)^2}
 +\frac{6[V(x)V'(x)^2+V(x)^2V''(x)]^2}{p^2V(x)^4V'(x)^2}\\ \nonumber
 &+&\frac{3[V(x)V'(x)^3+4V(x)^2V'(x)V''(x)+V(x)^3V'''(x)]} {8p^3V(x)^4V'(x)^4}
 -\frac{3[V(x)V'(x)^3+4V(x)^2V'(x)V''(x)
 +V(x)^3V'''(x)]}{p^2V(x)^3V'(x)^3})\\ \nonumber
             &=&-\frac{3}{16p^3x^3(1+x^2)^2}[2-5\sqrt{1+x^2}+4px(5+3x^2+195x^4+33x^5-84x^8-48x^{10})
 +x^2(-20+27\sqrt{1+x^2} \\ \nonumber
 &+&x^2(-106-21\sqrt{1+x^2}+x^2(-160+92x^4+90x^6+24x^8+17\sqrt{1+x^2}
 +x^2(-50+6\sqrt{1+x^2}))))]
 \end{eqnarray}
 In Fig.2, the running of spectral index is plotted against the parameter $p$ and it has been found that value of $\alpha_{s}$ is within the observational bound predicted by WMAP9 and PLANCK for a wide range of $p$.

 \begin{figure}[htbp]
 \begin{center}
\includegraphics[width=8cm]{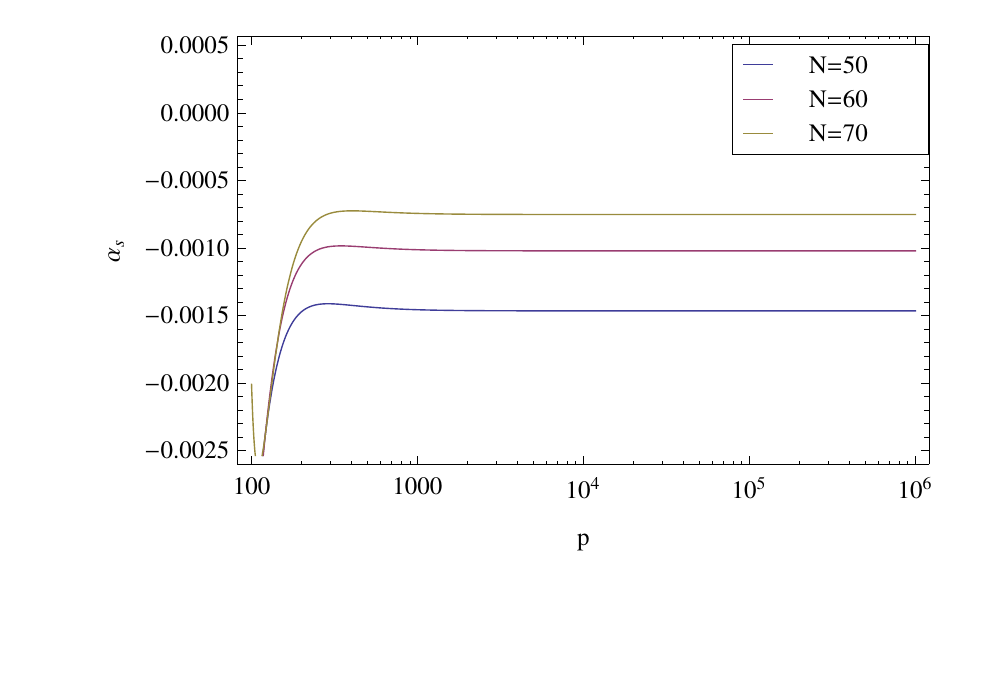}\\
\caption{Variation of the running of spectral index with respect to $p$ for N=70(brown),60(red) and 50(blue) respectively}\label{alpha}
\end{center}
\end{figure}

 \begin{flushleft}
\bf{(c) Power Spectrum : }
\end{flushleft}
 The power spectrum in terms of model parameters can be expressed as
 \begin{eqnarray}
 P &=&\frac{1}{4\pi^{2}}\left[\frac{1}{\sigma d}\frac{V^{6}}{(V,_{T})^{4}}\right]^{\frac{1}{4}}\exp(-2\bar{\zeta}(T))\\ \nonumber
  &=&\frac{\exp[\frac{-3\log[V(x)]}{8p}
-\frac{3\log[V'(x)]}{8p}-\frac{3}{32pV(x)}]
V(x)^{\frac{13}{4}}V'(x)^4
[\frac{fp^2V(x)^2}{V'(x)^4}]^{\frac{1}{4}}}
{4\pi^2}\\ \nonumber
    &=&\frac{1}{4\pi^2}\exp\left[\frac{-3\sqrt{1+x^2}}{32px}-\frac{3\log(\frac{x}{\sqrt{x^2+1}})}{8p}
-\frac{3\log(\frac{-x^2}{(1+x^2)^{\frac{3}{2}}}
+\frac{1}{\sqrt{x^2+1}})}{8p}\right]
\left[\frac{x}{\sqrt{x^2+1}}\right]^{(\frac{13}{4})}
\left[\frac{fp^2x^2}{(1+x^2)
(\frac{-x^2}{(1+x^2)^{\frac{3}{2}}}+ \frac{1}{\sqrt{x^2+1}})^4}\right]^{\frac{1}{4}} \\ \nonumber
 &\times &\left[\frac{1}{\sqrt{x^2+1}}
-\frac{x^2}{(1+x^2)^{\frac{3}{2}}}\right]^4
  \end{eqnarray}
 where, we have defined a dimensionless parameter $f$ as $f=\frac{16~L^4}{\sigma}$. The value of the parameter $f$ can be found out from the observational constraint on power spectrum as predicted by COBE normalization condition i.e $P=2\times10^{-9}$. The parameter $f$ can also be expressed in terms of $p$ and $N$ and hence it is not a new parameter in our model.

\begin{flushleft}
\bf{(d) Tensor to scalar ratio : }
\end{flushleft}

Another observable, the ratio of tensor spectrum to the scalar one in terms of model parameters reads as :

\begin{eqnarray}
R&=&\frac{2}{3M_{p}^{4}}\left[\frac{\sigma d [V,_{T}]^{4}}{V^{2}}\right]^{\frac{1}{4}}
\exp(2\bar{\zeta}(T))\left[\coth\left[\frac{k}{2T_{\gamma}}\right]\right]_{k=k_{0}} \\ \nonumber
  &=&\frac{32d\left[\frac{p^2 V(x)^2 V'(x)^4}{f}\right]^{\frac{1}{4}}
             \exp\left[\frac{3\log[V(x)]}{8p}+\frac{3\log[V'(x)]}{8p}+\frac{3}{32pV(x)}\right]}{3V(x)^{\frac{13}{4}V'(x)^4}} \\ \nonumber
  &=&\frac{32d\exp\left[\frac{3\sqrt{1+x^2}}{32px}+\frac{3\log(\frac{x}{\sqrt{x^2+1}})}{8p}
            +\frac{3\log(\frac{-x^2}{(1+x^2)^{\frac{3}{2}}}
               +\frac{1}{\sqrt{x^2+1}})}{8p}\right]
                 \left[\frac{p^2x^2[\frac{1}{\sqrt{x^2+1}}
                      -\frac{x^2}{(1+x^2)^{\frac{3}{2}}}]^4}{f(x^2+1)}\right]^{\frac{1}{4}}}
                        {3\left[\frac{1}{\sqrt{x^2+1}}-\frac{x^2}{(1+x^2)^{\frac{3}{2}}}\right]^4\left(\frac{x}{\sqrt{x^2+1}}\right)^{\frac{13}{4}}}
\end{eqnarray}

$R$ can be evaluated by putting the observational value of $d$ which is a combination of the radiation temperature  $T_{\gamma}$ and Stefan's constant $\sigma$  and $k_{0}$. From observation, using the fact that $T_\gamma \cong 0.24\times 10^{16} GeV $, $\sigma=1$ and $k_0=0.002Mpc^{-1}$ , $R$ can be measured in terms of parameters $p$ and $N$. In Fig.3 , $R$ is plotted against $p$ for three different values of $N$ and it has been found that for each value of $N$, value of $R$ lies well within the bound predicted by recent observation (both WMAP9 and PLANCK).

\begin{figure}[httb]
  \centering
  \includegraphics[width=8cm]{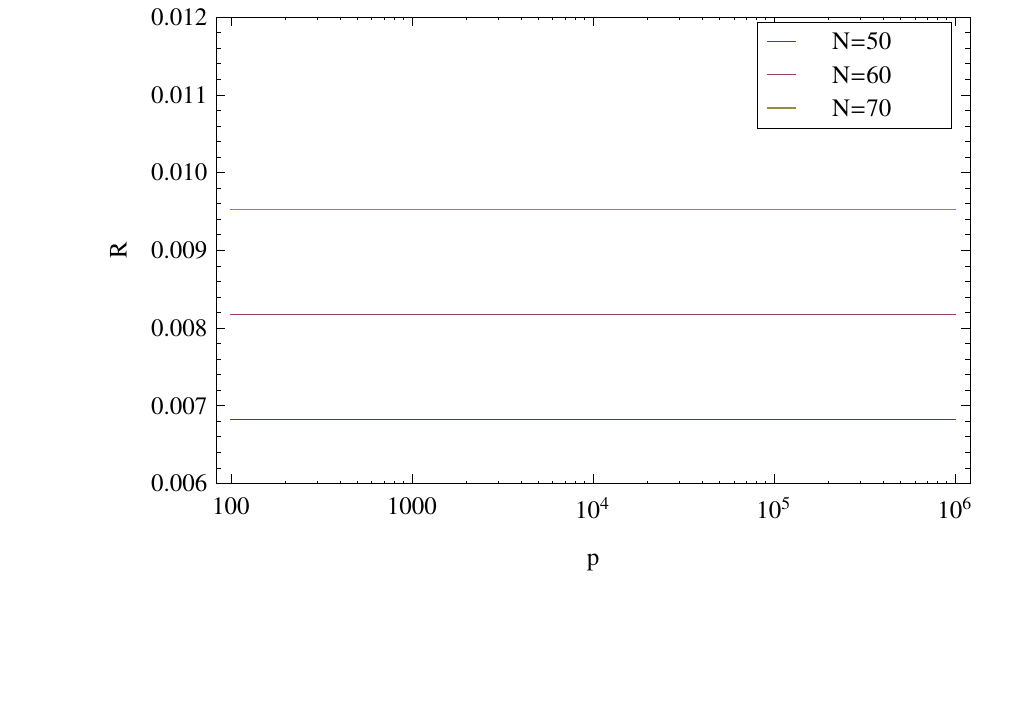}\\
  \caption{Variation of $R$ with the parameter $p$ for N=70(brown),60(red) and 50(blue) respectively }\label{ar}
\end{figure}

\subsubsection{Non-Gaussianity}

Non-gaussian statistics provides a powerful way to observationally discriminate between different mechanisms of generating curvature perturbation.Since an warm inflationary scenario can be viewed as multifield inflationary scenario,it may produce large non-gaussuanities. In order to study these non-gaussiani effects,we need to obtain the three point correlation function  of the density perturbation or the bispectrum \cite{GB}. The bi-spectrum is expressed in terms of the $f_{NL}$ parameter, the value of which predicts whether non-gaussianity is arising from the inflationary model or not.

In an warm inflationary scenario, the $f_{NL}$ parameter can be expressed in terms of the potential $V(x)$ as \cite{Kamali}
\begin{eqnarray}
f_{NL}&=&-\frac{5}{3}\left(\frac{\dot{T}}{H}\right)\left(\frac{1}{H}\log\left(\frac{k_{f}}{H}\right)\right)\left[\frac{V,_{TTT}}{\Gamma}+\frac{(2k_{f})^{2}V,_{T}}{\Gamma}\right]\\
&=&\frac{5 \log[\frac{\sqrt{3}V'(x)}{2\sqrt{d}}][\frac{[V(x)V'(x)^3+4V(x)^2V'(x)V''(x)
 +V(x)^3V'''(x)]}{V(x)^{\frac{3}{2}}V'(x)^2}+2p\sqrt{V(x)}V'(x)]}{3p^2V'(x)V(x)^{\frac{3}{2}}}.
\end{eqnarray}
In this case, $V(x)=\frac{x}{\sqrt{x^2+1}}$ and hence in terms of model parameters -
\begin{equation}
f_{NL}=\frac{5\log\left[\frac{\sqrt{3}}{2\sqrt{d}(1+x^2)^{\frac{3}{2}}}\right][1-14x^2-3x^4+12x^6+2px\sqrt{x^2+1}]}{3p^2x^2}
\end{equation}
Like the other observables, $f_{NL}$ can also be expressed in terms of the parameter $p$ and $N$.

In Fig.4, $f_{NL}$ parameter is plotted against the parameter $p$ for $N=50,60,70$ respectively. It can be seen that value of this parameter is well within the bound predicted by recent observations (both WMAP9 and PLANCK)for a large range of $p$.

\begin{figure}[httb]
  \centering
  \includegraphics[width=8cm]{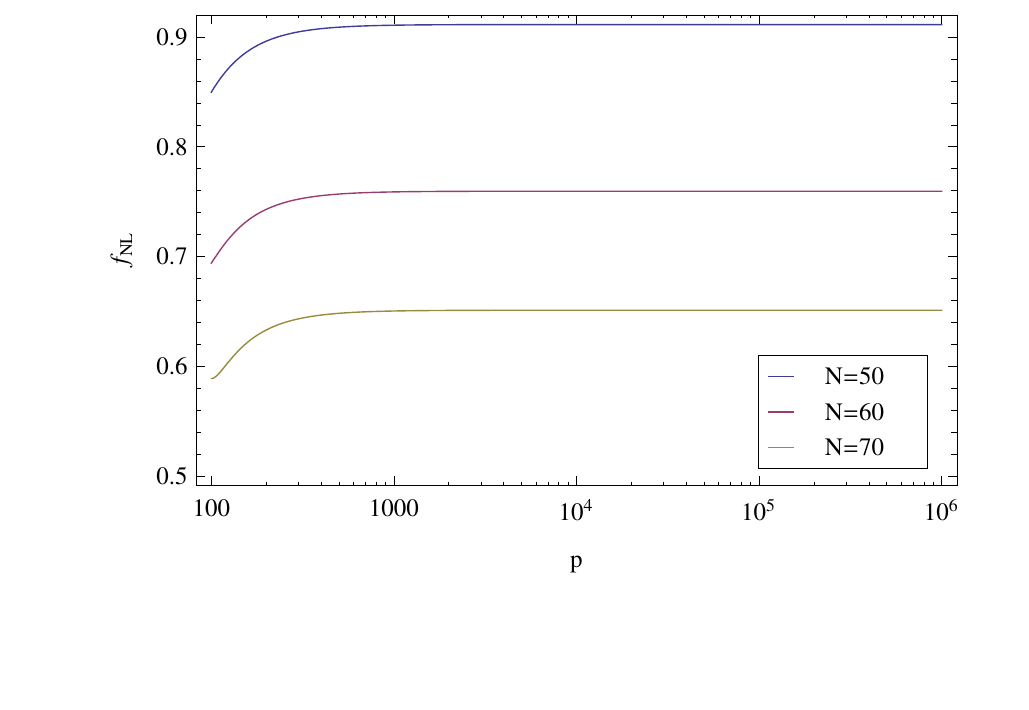}\\
  \caption{Variation of $f_{NL}$ with the parameter $p$ for N=70(brown),60(red) and 50(blue) respectively}\label{non-gauss}
\end{figure}

\subsubsection{Constraint on number of $D5$ branes}
  A constraint on the number of D5 branes to realize this model of inflation can be obtained as follows. From the whole analysis, we can see that all the observables lie well within the limit as predicted by recent observation for $p\geqslant 86$ {
   Now at the end point of inflation marked by $\varepsilon =1$ , we have
\begin{eqnarray}
\frac{1}{2V(x)}=1\\
\Rightarrow\frac{1}{2p\frac{x}{\sqrt{x^2+1}}}=1\\ \nonumber
\Rightarrow~x=\frac{1}{\sqrt{4p^2-1}}\\
\end{eqnarray}
As $x=\frac{R}{L}$ and $L=\sqrt{k~g_{s}l_{s}^2}$, so, we have,
\begin{equation}
g_{s}=\frac{(4p^2-1)R^2}{k~l_{s}^2}
\end{equation}
Now, using the conditions $R<\sqrt{k~g_{s}l_{s}^2}$ and $g_{s}<<1$, we arrive at the expression
\begin{equation}
k>(R/{l_s})^2~(4p^2-1)
\end{equation}

Now, using $p=100$, the number of branes required is of the order of $k>10^4$ and as we increase $p$, this requirment also increases. It may be interesting to note here that for a similar inflationary picture in absence of a radiation bath \cite{Panda2}, the constraint on the number of branes were obtained as $k g_s (l_s M_P)^2\approxeq 10^{10}$ which roughly leads to a impression that $k\geqslant 10^{10}$.

\section {Warm inflationary scenario driven by motion of D3 brane in the background of stack of static D5 branes with  $\Gamma = \frac{C_{T}T_{\gamma}^{3}}{T^2}$}

The dissipation coefficient $\Gamma$ has been computed in low temperature models \cite{Moss1,Berera10} from first principles in quantum field theory. Recently numerically dissipation co-efficient has been computed in supersymmetric models which have multiplets of heavy and light fields in addition to the inflaton field \cite{Berera8}. The expression for dissipation coefficient has been seen to have the form $\Gamma = \frac{C_{T} T_{\gamma}^{3}}{T^2}$ where, $C_{T}$ is a constant which depends on the coupling of the scalar field in the light and heavy mass sectors and its value changes in different temperature regimes.
Though for cosmological use, the space time has to be de-sitter which does not admit supersymmetry, we use this form of $\Gamma$ to see its implication in our analysis.

As mentioned in section II, the potential of the tachyonic scalar field which arises due to the motion of a D3 brane in the background of a stack of D5 brane expresses as
\begin{equation}
V(T)= \tau_{3} V(x)=\tau_{3}\frac{x}{\sqrt{x^2+1}},
\end{equation}
where $\tau_3$ is the tension on the branes and $x=\frac{R}{L}$. $R$ is the distance between the moving $D3$ and stack of $D5$ branes and  $L=\sqrt{k~g_{s}l_{s}^2}$ , $g_s$being the string coupling and $l_s$ being the string length scale.

Also, we have seen earlier that in terms of the dimensionless parameter $x$  the tachyonic field $T$ can be expressed as :
\begin{equation}
T=L \left(\sqrt{1+x^2}+\frac{1}{2} \log  \left( \frac{\sqrt{1+x^2}-1}{\sqrt{1+x^2}+1} \right)\right)
\end{equation}

To proceed further, we see that the dimensionless parameter $r$ in this case can be expressed as $r=\frac{f}{3HVT^2}$ where, $f=C_{T}~T_{\gamma}^3$.

And, with this form of $r$, the slow roll parameters in this scenario can be defined as
 \begin{eqnarray}
\varepsilon ~&=&~-\frac{\dot{H}}{H^2}=\frac{\sqrt{3}M_{P}~T^2~ V,_{T}^2}{2 ~f~ V^{\frac{3}{2}}}\\
&=& \frac{\sqrt{3~}M_{P}~ t~ \sqrt{V(x)} V'(x)^2  \left(\sqrt{1+x^2}+\frac{1}{2} \log  \left( \frac{\sqrt{1+x^2}-1}{\sqrt{1+x^2}+1}\right) \right)^2}{2} \\
\eta~&=&-\frac{\ddot{H}}{H \dot{H}}=\frac{\sqrt{3}~M_{p}~T^2}{f}\left(\frac{V,_{TT}}{\sqrt{V}}-\frac{V,_{T}^2}{2V^{\frac{3}{2}}}\right)\\
&=&\frac{\sqrt{3}~M_{P}~ t}{4}\left(\sqrt{1+x^2}+\frac{1}{2} \log  \left( \frac{\sqrt{1+x^2}-1}{\sqrt{1+x^2}+1} \right)\right)^2 \left(\sqrt{V(x)}V'(x)^2+ 2V(x)^{\frac{3}{2}}V''(x)\right) \\
\end{eqnarray}
where, $t$ is a parameter introduced as $t= \frac{\sqrt{\tau_{3}}}{f}$.

Also, the amount of inflation i.e, the number of e-foldings $N$ between to times $t_i$ and $t_f$ can be expressed as
\begin{eqnarray}
N(T)&=&{{\int}_{t_i}}^{t_f} H dt \\
&=&-\int_{T_{i}}^{T_{f}}\frac{f H}{V,_{T} T^{2}}~dT\\
&=& - \frac{1}{\sqrt{3}~t~M_{P}} \int_{x_{i}}^{x_{f}}\frac{dx}{V'(x) V(x)^{\frac{3}{2}} \left(\sqrt{1+x^2}+\frac{1}{2} \log  \left( \frac{\sqrt{1+x^2}-1}{\sqrt{1+x^2}+1} \right)\right)^2}\\
\end{eqnarray}

In this case, the end of the inflation can be designated as $\eta =1$ as $\eta$ approaches to 1 faster than $\varepsilon$ for a wide choice of parameter $t$.

Now for the perturbation analysis, both scalar and the tensor perturbations are considered and correspondingly, all the observables related to this perturbations can be computed whose values decide the validity of the warm inflationary model.

As the high dissipative regime is considered, so, the expression of density perturbation turns to be

\begin{equation}
\delta^{2}_{H}= \frac{\sqrt{3}}{75~\pi^{2}}\exp \left[-2 \bar{\zeta}(T)\right] \left[\left(\frac{1}{\varepsilon}\right)^3 \left(\frac{9 M_{P}^{2} T^2}{C_{T} r^{\frac{1}{2}}}\right)\right]^{\frac{1}{3}}
\end{equation}
where, the $\bar{\zeta}(T)$ parameter in high dissipative regime can be expressed as
\begin{eqnarray}
\bar{\zeta}(T)~&=&~-\int \left(\frac{1}{3Hr}\left(\frac{\Gamma}{V}\right),_{T}+\frac{9}{8}\left(1-\frac{(\ln(\Gamma)),_{T}(\ln(V)),_{T}}{36 H^{2}r}\right)(\ln(V)),_{T}\right)~ dT\\
~&=&~ \int \left(\frac{f~ V,_{T}}{3Hr~V^{2}~ T^2}+\frac{2~f}{3Hr~V~T^3}-\frac{9 V,_{T}}{8V}-\frac{V,_{T}^2}{16 V^2~r~ H^2~ T} \right)~ dT \\
\end{eqnarray}

With the value of $\bar{\zeta}(T)$, all the observables can be computed:

[1] Spectral Index:
\begin{eqnarray}
n_{s}-1 &=& \frac{d(\ln \delta_{H}^2)}{d(\ln k)}\\
&=& \frac{M_{p}^2 V,_{T}}{r ~V^2}\left(-2\zeta(T),_{T}+\frac{2}{3~T}-\frac{r,_{T}}{6~r}-\frac{\varepsilon,_{T}}{\varepsilon}   \right)\\
\end{eqnarray}

[2] Running of spectral index:
\begin{eqnarray}
\alpha_{s}~&=&~ \frac{d n_{s}}{d \ln k}\\
~&=&~ M_{P}^4 [ -\frac{2 V,_{T} V,_{TT} \zeta(T),_{T}}{r^2 V^4}-\frac{2 V,_{T}^2 \zeta,_{TT}}{r^2 V^4}+\frac{2 \zeta(T),_{T} r,_{T} V,_{T}^2}{r^3V^4}+\frac{4 V,_{T}^3 \zeta(T),_{T}}{r^2 V^5} +\frac{2 V,_{T} V,_{TT}}{3 T r^2 V^4}-\frac{2 V,_{T}^2 }{3 T r^2 V^4}-\frac{2V,_{T}^2 r,_{T}}{3 T r^3 V^4}\\ \nonumber
&-& \frac{4 V,_{T}^3}{3 T r^2 V^5}- \frac{V,_{T} V,_{TT} r,_{T}}{6 r^3 V^{4}}- \frac{V,_{T}^2 r,_{TT}}{6 r^3 V^{4}}+ \frac{V,_{T}^{2} r,_{T}^2}{3 r^4 V^4}+\frac{V,_{T}^3 r,_{T}}{3 r^3 V^5}- \frac{V,_{TT} V,_{T} \varepsilon,_{T}}{ r^2 \varepsilon V^4}- \frac{V,_{T}^2 \varepsilon,_{TT}}{r^2 \varepsilon V^4}+ \frac{V,_{T}^2 \varepsilon,_{T} r,_{T}}{r^3 \varepsilon V^4}\\ \nonumber
&+& \frac{V,_{T}^{2} \varepsilon,_{T}^{2}}{r^2 \varepsilon^{2} V^4}+\frac{2 \varepsilon,_{T} V_{T}^3}{r^2 \varepsilon V^5}  ]
\end{eqnarray}

[3] Power spectrum:
\begin{eqnarray}
P ~&=&~ \frac{25}{4} \delta_{H}^{2}\\
~&=&~ \frac{1}{4 \sqrt{3}\pi^{2}}\left (\frac{9 M_{P}^{2} T^2}{\sqrt{r} \varepsilon^3 C_{T}} \right)^{\frac{1}{3}} \exp[-2 \bar{\zeta}(T)]
\end{eqnarray}

[4] Tensor-scalar ratio:
\begin{eqnarray}
R ~&=&~ \left(\frac{A_{g}^2}{P} \right)_{k=k_{0}}\\
~&=&~ \frac{2 \sqrt{3}H^2 \coth \left(\frac{k}{2T_{\gamma}} \right)_{k=k_{0}}}{M_{P}^2}\exp[2 \bar{\zeta}(T)]\left(\frac{\sqrt{r}\varepsilon^3 C_{T}}{9 M_{P}^2 T^2} \right)^{\frac{1}{3}}\\
\end{eqnarray}

All the observables can be expressed in terms of $t$, $M_{P}$ and $x$ and the values of the observables are computed in the table below for different values of $t$ and $x$.
\pagebreak

\begin{table}
\renewcommand{\arraystretch}{1.5}
\caption{Table showing the values of the observables for different values of $t$,$x$ and $x_{f}$:}
\label{tab:example}
\centering
\begin{tabular}{|c|c|c|c|c|c|} \hline
    t   &   $x_{f}$   &   $x$   &   $n_{s}$   &   $\alpha$   &   R  \\
\hline
    60   &   0.39351229   &   81.50046   &   1.000091   &   -0.02245   &   $1.134\times10^{-9}$\\
\hline
    70   &   0.3954005   &   91.326678   &   1.000079   &   -0.02215   & $1.377\times 10^{-9}$\\
\hline
    80   &   0.39685695   &   94.1879768   &   1.000091   &   -0.02875   & $1.744 \times 10^{-9}$\\
\hline
    90   &   0.3980152   &   95.222316   &   1.00011   &   -0.03869   & $2,183 \times 10^{-9}$\\
\hline
    $10^{2}$   &   0.39895865   &   98.0457   &   1.00012   &   -0.04673   & $2.617 \times 10^{-9}$\\
\hline
    $10^{3}$   &   0.40724489   &   445.51776   &   1.000028   &   -0.010006   & $5.76 \times 10^{-8}$\\
\hline
\end{tabular}
\end{table}

From the table, it can be seen that for a wide range of parameter $t$, the values of the observables lie well within the bound predicted by recent observations like WMAP9.

\section{Conclusion}
In this piece of work, the warm inflationary scenario of a BPS D3 brane moving in the background of k coincident D5 branes has been analyzed. The unstable tachyonic field which arises from the motion of D3 brane in that background is the source of inflation in this picture.The whole process is considered in presence of a radiation bath. In such warm inflation,   the dissipative effects play important role and hence the whole dynamics is analyzed in terms of the dissipative parameter. The slow-roll parameters and the cosmological observables are computed by considering the fact that the dissipative parameter is a function of the tachyonic field. The number of e-foldings $N$, the spectral index $n_{s}$, running of spectral index $\alpha_{s}$ and the tensor-scalar ratio $R$ is evaluated for the general tachyonic potential in terms our model parameters. We have done the analysis for the three cases $N=50$, $N=60$ and $N=70$ separately and found that, for each case, all the observables lie well within the bound predicted by the recent observation WMAP9 and PLANCK for a wide range of the model parameters. The model predicts that by a suitable choice of parameter $p$, one can get a constraint on the minimum number of $D5$ branes  required to realize such a model. It is observed that for $p=100$, we need a minimum of $10^4$ branes to drive inflation.  Also, it is seen that non-gaussian effects can arise in this model due to the self-interaction of the inflaton field. The non-Gaussian effects of such inflationary mechanism  is analyzed by measuring the bispectrum of the gravitational field fluctuations generated during the warm inflation in strong dissipative regime.The bispectrum of the inflaton is expressed in terms of the parameter $f_{NL}$ and it can be seen that the value of $f_{NL}$ parameter lies well within observed limits for a wide range of the model parameters.
In addition, the scenario has been analyzed with a specific form of dissipation parameter. We observe that the model still remains a single parameter model. The range of parameter for which the cosmological observables are within WMAP9 (1$\sigma$) is wider in this case.Though the spectral index is always blue which does not conform with PLANCK observations.

\begin{center}
Acknowledgements
\end{center}
 Financial support from DAE Project grant 2009/24/35 (BRNS) is acknowledged. Hospitality provided by HRI, Allahabad and IUCAA, Pune during preparation of the manuscript is highly appreciated. We thank Prof A Berera and Prof S Panda for valuable discussions.


\end{document}